\documentclass[10pt,twoside]{article}
\usepackage{Latex-document}
\almvol{00}
\almttone{Frontiers of Science Awards}
\almtttwo{for Math/Phys/ISE}
\firstpage{1}
\usepackage[english]{babel}
\usepackage[utf8]{inputenc}
\usepackage{multicol}
\usepackage{tabu}
\usepackage{amsmath}
\usepackage{amssymb}
\usepackage{tikz}
\usepackage{lipsum}
\usepackage{braket}
\usepackage{graphicx}
\usepackage{esint}
\usepackage{epstopdf}
\usepackage{color}
\usepackage{hyperref}
\usepackage{mathtools}
\usepackage{stackengine}
\usepackage{appendix}
\usepackage{leftindex}
\usepackage{soul}
\usepackage{multirow}
\usepackage{placeins}
\usepackage{xspace}
\usepackage{nicefrac}
\usepackage{lineno}
\usepackage{tikz-feynman}

\numberwithin{equation}{section}

\newcommand{\ttbar}{\ensuremath{t \bar{t}}\xspace}
\newcommand{\Mtt}{\ensuremath{M_{t\bar{t}}}\xspace}
\newcommand{\trans}{\ensuremath{^{\textrm{T}}}\xspace}

\begin{document}

\markboth{\hfill{\rm Yoav Afik and Juan Ram\'on Muñoz de Nova} \hfill}{\hfill {\rm Quantum Information in High-Energy Physics: a top subject \hfill}}

\title{Quantum Information in High-Energy Physics: a top subject}

\author{Yoav Afik and Juan Ram\'on Mu\~noz de Nova}
\address{Enrico Fermi Institute, University of Chicago\\ 
Chicago, Illinois 60637, USA\\
\email{yoavafik@gmail.com}}

\address{Instituto de Estructura de la Materia, IEM-CSIC\\ 
Serrano, 123 E-28006 Madrid, Spain\\
\email{jr.denova@csic.es}}

\begin{abstract}
As a work selected for the Frontiers of Science Award, we present in the Proceedings of the ICBS an overview of how the techniques from the field of Quantum Information (QI) can be implemented in High-Energy Physics (HEP), including also some novel insights. 
We focus on the paradigmatic case of the top quark, a drosophila of a relativistic qubit. 
We study the presence of entanglement, perhaps the most genuine feature of Quantum Mechanics, in top-antitop quark production in high-energy colliders such as the LHC. 
We also analyze our experimental proposal of a quantum tomography protocol for the top-antitop quark pair, eventually leading to the entanglement observations by the ATLAS and CMS collaborations, which constituted the highest-energy observations of entanglement ever.
We discuss how the concepts and techniques developed in our work have effectively spawned an already vast and rich field of QI in HEP.
\end{abstract}

\maketitle

\setcounter{tocdepth}{1}

\section{Introduction}\label{sec:Intro}

Quantum Information (QI) is currently one of the hottest fields of research in science, leveraging the fundamental principles of Quantum Mechanics to design completely novel devices able to process information and carry out computations well beyond the capabilities of their classical counterparts. Within this field, entanglement, perhaps the most genuine quantum feature, stands out as an essential ingredient of the quantum advantage.

On the other hand, High-Energy Physics (HEP) studies the fundamental particles and interactions present in nature, establishing the current frontier of our understanding of the universe. 
The experimental HEP program is carried out in high-energy colliders, such as the Large Hadron Collider (LHC), smashing particle beams at energies about $\sim10^{12}$ times higher than conventional laboratory setups. 
As a result, high-energy colliders provide a unique fundamental environment of a genuine relativistic nature where to study the foundational quantum concepts underlying QI, e.g., entanglement.

A number of works in the literature have addressed entanglement across very diverse scenarios in relativistic quantum field theories, ranging from the proton composition~\cite{Baker:2017wtt} to Hawking radiation in black holes~\cite{Page1993}. 
However, in comparison, very little attention has been devoted to experimental HEP studies of entanglement and other fundamental quantum features (see Refs.~\cite{Go:2007ww,Formaggio:2016cuh} for some exceptions involving mesons and neutrinos). 
For this purpose, it is crucial to provide a precise dictionary that allows to translate QI concepts into HEP observables and vice versa, as well as to identify feasible HEP platforms for the implementation of these ideas. 

This is where our awarded work emerged~\cite{Afik2021} (and its subsequent generalization~\cite{Afik2022}), effectively bridging the gap between QI theory and experimental HEP. 
Our original motivation was the adaptation of the work of one of us~\cite{deNova2014,deNova2015}, where the violation of Cauchy-Schwarz inequalities and other quantum-optics techniques~\cite{Walls2008} were proposed for the experimental characterization of the Hawking effect, to collider setups. 
The underlying idea was that condensed matter, quantum optics, and particle physics share the common framework of quantum field theory, which should thus allow for a continuous conceptual flow between them.

Here, we present an overview of  Refs.~\cite{Afik2021,Afik2022}, together with some novel insights gained in retrospect.
We first address the translation of QI concepts into the HEP language, focusing on the paradigmatic case of the top quark, which is a \textit{drosophila} of a relativistic qubit due to its unique polarimetric properties~\cite{Bernreuther1998}. 
We study entanglement in top-antitop quark ($t\bar{t}$) production at the LHC, and propose an experimental quantum tomography protocol for the \ttbar quantum state. 
We review how our experimental proposal was implemented by the ATLAS and CMS collaborations~\cite{CMS:2025kzt,ATLAS2026}, resulting in the highest-energy observations of entanglement ever.
Finally, we discuss the impact of our work and how it has critically contributed to the development of the field of QI in HEP.

\section{General formalism}\label{sec:Formalism}

\subsection{Quantum correlations}\label{subsec:Quantum}

In Quantum Mechanics, pure quantum states are described by a unit vector\footnote{Technically, a unit ray $\{e^{i\alpha}\ket{\psi}\}_{\alpha\in[0,2\pi)}$, $\braket{\psi|\psi}=1$.} $\ket{\psi}$ in a Hilbert space $\mathcal{H}$ of dimension $N$. 
Given a certain orthonormal basis $\{\ket{i}\}^N_{i=1}$ of the Hilbert space, any quantum state is expanded as
\begin{equation}\label{eq:PureState}
    \ket{\psi}=\sum^N_{i=1} c_i\ket{i}.
\end{equation}
Observables are described by Hermitian operators $O$ in $\mathcal{H}$, and their expectation values are simply computed as $\braket{O}=\braket{\psi|O|\psi}$.

A more general description of a quantum state is given by its density matrix $\rho$, a non-negative Hermitian operator with unit trace, $\textrm{Tr}[\rho]=1$. 
The density matrix is expanded in terms of its matrix elements as
\begin{equation}
    \rho=\sum^N_{i,j=1}\rho_{ij}\ket{i}\bra{j}.
\end{equation}
In particular, a pure state is described by a rank-$1$ projector $\rho=\ket{\psi}\bra{\psi}$. 
A deeper conceptual insight is obtained by expressing $\rho$ in terms of its eigenstates:
\begin{equation}
    \rho=\sum^N_{n=1} p_n\ket{\psi_n}\bra{\psi_n},~0\leq p_n\leq 1,~\textrm{Tr}[\rho]=\sum^N_{n=1} p_n=1.
\end{equation}
We can then understand $\rho$ as an \textit{incoherent} statistical mixture of the pure states $\ket{\psi_n}$ with a probability $p_n$. 
Expectation values are now computed by taking the product trace,
\begin{equation}
    \braket{O}=\textrm{Tr}[O\rho]=\sum^N_{n=1} p_n\braket{\psi_n|O|\psi_n},
\end{equation}
which is simply the average of the expectation values in each state $\ket{\psi_n}$. 
In this way, density matrices provide a link between quantum and classical descriptions.

The simplest non-trivial quantum-mechanical system is a qubit, i.e., a two-level quantum system, such as the spin-1/2 of a particle\footnote{In general, any qubit can be always regarded as representing the spin-$1/2$ of a fictitious particle, denoted in that case as pseudospin.}. A conventional basis of the Hilbert space is given by the eigenstates of the third component of the spin, $\sigma^3\ket{\pm}=\pm \ket{\pm}$, with $\sigma^i$, $i=1,2,3$, the operators described by the usual Pauli matrices. 
Any quantum state in this Hilbert space can be then written as 
\begin{equation}
    \ket{\psi}=\cos\frac{\theta}{2}\ket{+}+e^{i\phi}\sin\frac{\theta}{2}\ket{-}\equiv \ket{\mathbf{\hat{n}}},~\theta\in[0,\pi],~\phi\in[0,2\pi),
\end{equation}
$\mathbf{\hat{n}}=[\sin \theta \cos\phi,\sin \theta \sin\phi,\cos \theta]$ being a unit vector, spanning the surface of the Bloch sphere. 
This implies that any qubit state can be always thought as an eigenstate of the spin projection along a certain direction $\mathbf{\hat{n}}$, $(\boldsymbol{\sigma}\cdot\mathbf{\hat{n}})\ket{\pm \mathbf{\hat{n}}}=\pm \ket{\pm \mathbf{\hat{n}}}$.

On the other hand, the density matrix of a single qubit can be expanded as
\begin{equation}\label{eq:GeneralQubitQuantumState}
    \rho=\frac{I_2+\sum_{i}B_{i}\sigma^i}{2}=\frac{1+|\mathbf{B}|}{2}\ket{\mathbf{\hat{n}}}\bra{\mathbf{\hat{n}}}+\frac{1-|\mathbf{B}|}{2}\ket{\mathbf{-\hat{n}}}\bra{\mathbf{-\hat{n}}},
\end{equation}
where $I_n$ is the $n\times n$ identity matrix and $\mathbf{B}=|\mathbf{B}|\mathbf{\hat{n}}$ is the Bloch vector, whose components $B_{i}=\braket{\sigma^i}$ represent the spin polarization. 
Physical (i.e., non-negative) density matrices correspond to Bloch vectors inside the Bloch sphere, $|\mathbf{B}|\leq 1$.

An alternative description of $\rho$ is provided by the $P$-representation (the qubit counterpart of the celebrated Glauber-Sudarshan distribution~\cite{Walls2008}):
\begin{equation}\label{eq:PRepresentationQubit}
    \rho=\int \mathrm{d}\Omega~P(\mathbf{\hat{n}})\ket{\mathbf{\hat{n}}}\bra{\mathbf{\hat{n}}},~\int \mathrm{d}\Omega~P(\mathbf{\hat{n}})=1,
\end{equation}
with $\Omega$ the solid angle associated to the unit vector $\mathbf{\hat{n}}$. 
Notice that $P(\mathbf{\hat{n}})$ is a \textit{quasi-probability} distribution, since it does not describe disjoint states, $|\braket{\mathbf{\hat{n}}|\mathbf{\hat{n}}'}|^2=(1+\mathbf{\hat{n}}\cdot \mathbf{\hat{n}}')/2\neq 0$.
Nevertheless, $P(\mathbf{\hat{n}})$ can be always chosen as non-negative for a single qubit [e.g., Eq.~(\ref{eq:GeneralQubitQuantumState})]. 

More intriguing physics emerges in bipartite Hilbert spaces $\mathcal{H}=\mathcal{H}_A\otimes\mathcal{H}_B$ composed of subsystems $A,B$, to which we will refer as Alice and Bob for simplicity. 
Using a product basis $\ket{ij}\equiv\ket{i}\otimes \ket{j}$, any quantum state can be expressed as
\begin{equation}
   \ket{\psi}=\sum_{i,j}c_{ij}\ket{ij},~ 
\end{equation}
where hereafter we skip the sum limits for the sake of compactness. 
A remarkable consequence of the superposition principle is that, while the chosen basis is a tensor product, not all states factorize in this form. 
Specifically, product states
\begin{equation}
    \ket{\psi}=\ket{\phi}\otimes\ket{\chi}
\end{equation}
are said to be \textit{separable}, while non-separable states are said to be \textit{entangled}. 
By invoking the singular-value decomposition, it is immediate to prove that separability is equivalent to the coefficients $c_{ij}$ spanning a rank-$1$ matrix. 
Alternatively, a state is separable \textit{iff} the expectation value of all products of local observables $A\otimes B$ (describing local measurements of Alice and Bob) factorize, $\braket{A\otimes B}=\braket{A}\otimes \braket{B}$. 
Consequently, for pure states, entanglement is equivalent to the presence of correlations between Alice and Bob.

However, with the advent of the field of Quantum Information, the conceptual distinction between classical and genuinely quantum correlations became crucial, as the description of realistic experimental scenarios inevitably involves density matrices.
In particular, the notion of entanglement (i.e., non-separability) is translated into the density-matrix formalism by now defining separable states as those which can be written as a convex sum of product states,
\begin{equation}\label{eq:Separability}
\rho=\sum_n p_n \rho^{A}_{n}\otimes\rho^{B}_{n},~\quad \sum_n p_n=1,~\quad p_n\geq 0.
\end{equation}
Indeed, for general quantum states, a whole conceptual hierarchy of bipartite correlations exists, with a well-defined logical relation between them~\cite{Ollivier2001,Wiseman2007}:
\begin{equation*}
\textrm{Bell Nonlocality} \subset \textrm{Steering} \subset 
\textrm{Entanglement} \subset \textrm{Discord}\subset \textrm{Correlation}.
\end{equation*}
Here, ``\textrm{Correlation}'' denotes the presence of correlations between Alice and Bob, $\braket{A\otimes B}\neq \braket{A}\otimes \braket{B}$.
Notice that separability~(\ref{eq:Separability}) does not anymore imply $\braket{A\otimes B}=\braket{A}\otimes \braket{B}$, revealing the existence of classical correlations. 
Actually, the most basic form of quantum correlations is quantum discord~\cite{Ollivier2001}, while Bell nonlocality (the capability of a quantum state to violate a Bell inequality) represents the tighter correlation in this hierarchy. 
Steering sits between entanglement and Bell nonlocality, denoting the ability of Alice/Bob to manipulate the reduced state of Bob/Alice through local measurements~\cite{Wiseman2007}.
For pure states, correlation is equivalent to entanglement (as discussed above), while Gisin theorem~\cite{Gisin1991} shows that any entangled pure state violates a certain Bell inequality. 
Therefore, entanglement alone captures all possible forms of correlation in pure states, for which the whole hierarchy becomes equivalent, a fact that has strongly hindered its formulation and understanding\footnote{For instance, both discord and steering were precisely formulated only in the 21st century.}, inducing a lot of confusion among many physicists.

Nevertheless, entanglement is still widely regarded as the most genuine manifestation of quantum behavior, providing the sharpest classical/quantum boundary. To illustrate this idea, we consider the qubit case once more. For a two-qubit system, any density matrix $\rho$ can be decomposed as
\begin{equation}\label{eq:GeneralBipartiteStateRotations}
\rho=\frac{I_4+\sum_{i}\left(B^{+}_{i}\sigma^i\otimes I_2+B^{-}_{i}I_2\otimes\sigma^i \right)+\sum_{i,j}C_{ij}\sigma^{i}\otimes\sigma^{j}}{4},
\end{equation}
where the vectors $\mathbf{B}^{\pm}$ are the individual spin polarizations of Alice and Bob, $B^{+}_{i}=\braket{\sigma^i\otimes I_2},~B^{-}_{i}=\braket{I_2\otimes\sigma^i}$, and their spin correlations are encoded in the coefficients $C_{ij}=\braket{\sigma^{i}\otimes\sigma^{j}}$ of the matrix $\mathbf{C}$. 

An insightful way to visualize entanglement for two qubits emerges from the joint $P$-representation 
\begin{equation}\label{eq:PRepresentation2Qubit}   \rho=\int\mathrm{d}\Omega_A\mathrm{d}\Omega_B~P(\mathbf{\hat{n}}_A,\mathbf{\hat{n}}_B)\ket{\mathbf{\hat{n}}_A\mathbf{\hat{n}}_B}\bra{\mathbf{\hat{n}}_A\mathbf{\hat{n}}_B}, \quad \ket{\mathbf{\hat{n}}_A\mathbf{\hat{n}}_B}=\ket{\mathbf{\hat{n}}_A}\otimes\ket{\mathbf{\hat{n}}_B}.
\end{equation}
From Eqs.~(\ref{eq:GeneralQubitQuantumState}),~(\ref{eq:Separability}), it is seen that a state is separable \textit{iff} it admits a non-negative $P$-representation. 
This implies that expectation values for separable states can be evaluated in terms of  classical statistical averages. For instance,
\begin{equation}
    C_{ij}=\braket{\sigma^{i} \otimes\sigma^{j}}=\int\mathrm{d}\Omega_A\mathrm{d}\Omega_B~P(\mathbf{\hat{n}}_A,\mathbf{\hat{n}}_B)n^i_An^j_B.
\end{equation}
Thus, entanglement is equivalent to a negative-valued $P$-distribution, a genuine non-classical signature: if one insists on locally understanding an entangled system in terms of its components, then one needs to deal with counterintuitive negative probabilities. 
Conversely, one can use only well-defined non-negative distributions to characterize the system, but then at the price of working with a global nonlocal description of its components, which then become \textit{entangled} in a way that does not allow to describe them separately, something also beyond our classical intuition.

In order to signal the presence of entanglement, several theoretical criteria can be used (see Ref.~\cite{Horodecki2009} for a review on the topic). 
Perhaps the most common one is the Peres-Horodecki criterion, which simply states that, if $\rho$ is separable, the operator $\rho^{\rm{T_B}}$ resulting from transposing the matrix elements of $\rho$ in Bob subspace is also non-negative. 
Therefore, a negative-valued $\rho^{\rm{T_B}}$ implies that $\rho$ is entangled. 
In qubit-qubit and qubit-qutrit systems, the Peres-Horodecki criterion is also a necessary condition for entanglement. 

For two qubits, a more quantitative entanglement measurement is given by the concurrence:
\begin{equation}\label{eq:Concurrence}
\mathcal{C}[\rho]\equiv \max(0,\lambda_1-\lambda_2-\lambda_3-\lambda_4),
\end{equation}
$\lambda_i$ being the eigenvalues of the matrix $\sqrt{\sqrt{\rho}\tilde{\rho}\sqrt{\rho}}$ ordered in decreasing magnitude, $\tilde{\rho}=(\sigma^2\otimes\sigma^2)~\rho^*~(\sigma^2\otimes\sigma^2)$, and $\rho^*$ the complex conjugate of $\rho$ in the spin basis of $\sigma^3$. The concurrence satisfies $0\leq \mathcal{C}[\rho]\leq 1$, where a quantum state is entangled \textit{iff}  $\mathcal{C}[\rho]>0$. A maximally entangled ($\mathcal{C}[\rho]=1$) orthonormal basis of the Hilbert space is given by the four Bell states,
\begin{equation}
\label{eq:BellStates}
    \ket{\Phi^\pm}=\frac{\ket{++}\pm\ket{--}}{\sqrt{2}},~\ket{\Psi^\pm}=\frac{\ket{+-}\pm\ket{-+}}{\sqrt{2}}.
\end{equation}
In principle, the exact evaluation of the above entanglement criteria requires a complete knowledge of the density matrix. 
Experimentally, $\rho$ can be reconstructed through the measurement of the expectation value of a complete quorum of observables, a technique known as \textit{quantum tomography}. 
For instance, for a single qubit, it is enough to measure the spin polarization, yielding the Bloch vector $\mathbf{B}$.
For two qubits, one needs the $15$ parameters determining $\mathbf{B}^{\pm},\mathbf{C}$, obtained from the measurement of the spin polarizations and spin correlations. 

Nevertheless, simpler experimental criteria can be formulated using entanglement witnesses, defined as observables whose expectation values satisfy a given bound for all separable states, while this bound can be violated by entangled states, providing in this way sufficient conditions for entanglement. 
For instance, 
\begin{equation}\label{eq:DeltaQI}
    \Delta\equiv \frac{-C_{33}+|C_{11}+C_{22}|-1}{2}
\end{equation}
is an entanglement witness, since $\Delta\leq 0$ for all separable states. 
Hence, $\Delta>0$ is a sufficient condition for entanglement. 
Remarkably, both sign combinations can be written in terms of the fidelity with respect to a certain Bell state,
\begin{align}\label{eq:DeltaFidelity}
   \frac{-\textrm{Tr}[\mathbf{C}]-1}{2}>0 &\Longleftrightarrow \braket{\Psi^-|\rho|\Psi^-}>\frac{1}{2},\\
   \nonumber \frac{-C_{33}+C_{11}+C_{22}-1}{2}>0 &\Longleftrightarrow \braket{\Psi^+|\rho|\Psi^+}>\frac{1}{2}.
\end{align}
Indeed, all possible permutations in Eq.~(\ref{eq:DeltaQI}) yield an entanglement witness which can be similarly expressed in terms of the fidelity to one of the Bell states (\ref{eq:BellStates}).

\subsection{Quantum states in high-energy processes}\label{subsec:QuantumHEP}

The natural question now is: how can we translate the above ideas into the high-energy arena?
The simplest high-energy candidate for an entangled two-qubit system is a particle-antiparticle ($P\bar{P}$) pair, with $P$ a fundamental fermion (which thus has spin-$1/2$) of mass $m$, produced from some initial state $I$ through some gauge interaction that conserves energy and momentum:
\begin{equation}
    I\rightarrow P+\bar{P}.
\end{equation}

The $P\bar{P}$ kinematics is described in the center-of-mass (COM) frame by the invariant mass, $M$, and the $P$-flight direction, $\hat{k}$. In terms of the $P$ and $\bar{P}$ momenta, $k^{\mu}=(k^0,\mathbf{k})$ and $\bar{k}^{\mu}=(k^0,-\mathbf{k})$, with $\bar{k}^2=k^2\equiv k^{\mu} k_{\mu}=m^2$, they read as
\begin{equation}
M^2\equiv(k+\bar{k})^2=4(m^2+|\mathbf{k}|^2),~\quad\hat{k}=\frac{\mathbf{k}}{|\mathbf{k}|},~\quad\beta=\frac{|\mathbf{k}|}{k^0}=\sqrt{1-\frac{4m^2}{M^2}},
\end{equation}
$\beta$ being the particle COM velocity. Regarding the $P\bar{P}$ quantum state, for a certain production process with $(M,\hat{k})$, we define the production spin density matrix as
\begin{equation}\label{eq:ProductionSpinDensityMatrixT}
R^{I}_{\alpha\beta,\alpha'\beta'}(M,\hat{k})\equiv \frac{1}{N_\lambda}\sum_\lambda \mathcal{M}_{I\lambda\to M\hat{k}\alpha\beta}\mathcal{M}^*_{I\lambda\to M\hat{k}\alpha'\beta'}\,,
\end{equation}
where $\ket{I\lambda}$ denotes the initial state $I$ with internal degrees of freedom $\lambda$ (whose total number is $N_\lambda$, including spin and/or color), $\ket{M\hat{k}\alpha\beta}$ is the $P\bar{P}$ state with kinematics $(M,\hat{k})$ and spins $\alpha\beta$\footnote{The $P\bar{P}$ spins are evaluated in their respective rest
frames, where they are well defined.}, respectively, and $\mathcal{M}_{I\lambda\to M\hat{k}\alpha\beta}$ is the matrix element connecting them. The resulting $P\bar{P}$ quantum state reads 
\begin{equation}\label{eq:QuantumStateRSpin}
\rho^I=\frac{1}{\sigma^I}\sum_{\alpha\beta,\alpha'\beta'}\int \mathrm{d}\Omega~\frac{\mathrm{d}\sigma^I}{\mathrm{d}\Omega~}\rho^{I}_{\alpha\beta,\alpha'\beta'}(M,\hat{k})\frac{\ket{M\hat{k}\alpha\beta}\bra{M\hat{k}\alpha'\beta'}}{\braket{M\hat{k}|M\hat{k}}},~\quad\sigma^I=\int \mathrm{d}\Omega~\frac{\mathrm{d}\sigma^I}{\mathrm{d}\Omega~},
\end{equation}
with the actual spin density matrix $\rho^{I}(M,\hat{k})$ obtained from the normalization of $R^I(M,\hat{k})$, whose trace is in turn proportional to the COM differential cross section:
\begin{equation}
    \rho^{I}_{\alpha\beta,\alpha'\beta'}(M,\hat{k})=\frac{R^{I}_{\alpha\beta,\alpha'\beta'}(M,\hat{k})}{\mathrm{Tr}[R^I(M,\hat{k})]},\,\textrm{Tr}[R^I(M,\hat{k})]=\frac{1}{N_\lambda}\sum_{\alpha,\beta} \sum_\lambda |\mathcal{M}_{I\lambda\to M\hat{k}\alpha\beta}|^2\propto\frac{\mathrm{d}\sigma^I}{\mathrm{d}\Omega}.
\end{equation}
Thus, the $R$-matrix encodes all the information about the $P\bar{P}$ quantum state. The density-matrix description arises naturally in colliders due to the fact that i) the $P\bar{P}$ kinematics is measured event by event, projecting the quantum state in momentum and thereby resulting in a diagonal density matrix in the kinematical sector; and ii) one needs to average over the initial internal degrees of freedom [sum over $\lambda$ in Eq. (\ref{eq:ProductionSpinDensityMatrixT})] as these cannot be typically controlled in experiments.

\section{Entanglement in $t\bar{t}$ production via QCD}
\label{sec:entanglementfundamental}

The results of the previous section establish a dictionary between the high-energy amplitudes characterizing a certain $P\bar{P}$ production process and the resulting $P\bar{P}$ quantum state. 
As discussed in Sec.~\ref{sec:Experimental}, a particularly interesting realization of a relativistic two-qubit system is provided by a \ttbar pair.
Specifically, we study $t\bar{t}$ production via QCD processes using perturbation theory at leading-order (LO), since it provides analytical results and a neat picture of the underlying physics. Higher-order corrections are known to be small and do not alter the main results.

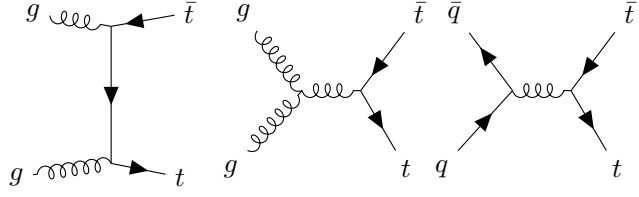
\begin{figure}[ht]
\centering
\resizebox{0.70\textwidth}{!}{ 
\begin{tikzpicture}
  \begin{feynman}

    \vertex (f1) {$g$};
    \vertex[right=1cm of f1] (f2);
    \vertex[right=1cm of f2] (f3);
    \vertex[below=6em of f3] (g1);
    \vertex[left=1cm of g1] (g2);
    \vertex[left=1.0cm of g2] (g3) {$g$};
    \vertex[below=3.0em of f2] (h1);
    \vertex[right=0.5cm of h1] (h2);
    \vertex[below=0.5em of f2] (i1);
    \vertex[above=0.5em of g2] (i2);
    \vertex[right=2.0cm of f1] (j1) {$\bar{t}$};
    \vertex[right=2.1cm of g3] (j2) {$t$};

    \vertex[right=2.7cm of f1] (k1) {$g$};
    \vertex[right=0.75cm of k1] (k2);
    \vertex[below=1.0cm of k2] (l1);
    \vertex[below=1.0cm of l1] (l2);
    \vertex[left=0.7cm of l2] (k3) {$g$};
    \vertex[right=0.75cm of l1] (m1);
    \vertex[right=2.25cm of k1] (n1){$\bar{t}$};
    \vertex[right=2.25cm of k3] (n2){$t$};

    \vertex[right=2.7cm of k1] (o1) {$\bar{q}$};
    \vertex[right=0.75cm of o1] (o2);
    \vertex[below=1.0cm of o2] (p1);
    \vertex[below=1.0cm of p1] (p2);
    \vertex[left=0.7cm of p2] (o3) {$q$};
    \vertex[right=0.75cm of p1] (q1);
    \vertex[right=2.25cm of o1] (r1){$\bar{t}$};
    \vertex[right=2.25cm of o3] (r2){$t$};

    \diagram* {
      {[edges=fermion]
      },

      (f1) -- [gluon] (i1),
      (i1) -- [fermion] (i2),
      (i2) -- [gluon] (g3),
      (i1) -- [anti fermion] (j1),
      (i2) -- [fermion] (j2),

      (k1) -- [gluon] (l1),
      (k3) -- [gluon] (l1),
      (l1) -- [gluon] (m1),
      (m1) -- [anti fermion] (n1),
      (m1) -- [fermion] (n2),

      (o1) -- [anti fermion] (p1),
      (o3) -- [fermion] (p1),
      (p1) -- [gluon] (q1),
      (q1) -- [anti fermion] (r1),
      (q1) -- [fermion] (r2),

    };

  \end{feynman}
\end{tikzpicture}
}
\caption{Representative Feynman diagrams for $t\bar{t}$ production via QCD.}
\label{fig:Scheme}
\end{figure}

At LO QCD, a $t\bar{t}$ is produced solely through two channels: quark-antiquark ($q \bar{q}$) annihilation or gluon-gluon ($gg$) fusion,
\begin{align}\label{eq:partonreactions}
    q+\bar{q}\rightarrow t+\bar{t},~\qquad g+g\rightarrow t+\bar{t}.
\end{align}
Representative Feynman diagrams for each process are shown in Fig.~\ref{fig:Scheme}. 

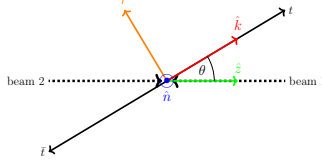
\begin{figure}[ht]
\centering

\resizebox{0.35\textwidth}{!}{ 
\begin{tikzpicture}

 \draw[<-,thick, dashed, line width=3.0] (-0.2,0)--(-5,0) node [left] {\Large beam 2}; 
 \draw[<-,thick, dashed, line width=3.0] (0.2,0)--(5,0) node [right] {\Large beam 1}; 
 \draw[->,thick, line width=2.0] (0,0)--(5,3) node [right] {\Large $t$}; 
 \draw[->,thick, line width=2.0] (0,0)--(-5,-3) node [left] {\Large $\bar t$}; 
 \draw[->,thick, draw=orange, line width=2.0] (0,0)--(-1.8,3) node [above, yshift=0.1cm] {\textcolor{orange}{\LARGE\bf$\hat r$}}; 

 \draw[->,thick, draw=red, line width=2.0] (0,0)--(3,1.8) node [above, yshift=0.2cm] {\textcolor{red}{\LARGE\bf$\hat k$}}; 

 \draw[->,thick, draw=green, line width=2.0] (0,0)--(3,0) node [above, yshift=0.2cm] {\textcolor{green}{\LARGE\bf$\hat z$}}; 

\node[blue] at (0,0) {\Huge \bf$\odot$} node [below, yshift=-0.3cm] {\textcolor{blue}{\LARGE\bf$\hat n$}};
\node[blue] at (0,0) {\LARGE \bf$\bullet$};

\draw[thick]
    (2,0) arc[start angle=0,end angle=31,radius=2cm];

\node at (1.5,0.45) {\LARGE \bf$\theta$};

\end{tikzpicture}
}
\caption{Schematic representation of the helicity basis.}
\label{fig:basis}
\end{figure}

The initial state $I=q\bar{q},gg$ gives rise to a $t\bar{t}$ spin quantum state described by the density matrix $\rho^{I}(M_{t\bar{t}},\hat{k})$, $M_{t\bar{t}}$ being the $t\bar{t}$ invariant mass and $\hat{k}$ the top flight direction in the \ttbar COM frame. 
Equivalently, we can work with $\rho^{I}(\beta,\hat{k})$, where $\beta$ is the top COM velocity.
The spin coefficients determining $\rho^{I}(\beta,\hat{k})$, Eq. (\ref{eq:GeneralBipartiteStateRotations}), are conventionally computed in the helicity basis, here defined in the COM frame as $\{\hat{k},\hat{r},\hat{n}\}$, with $\hat{r}=(\cos\Theta\hat{k}-\hat{p})/\sin\Theta$ and $\hat{n}=\hat{k}\times\hat{r}$, where $\hat{p}$ is the initial beam direction and $\cos\Theta=\hat{k}\cdot\hat{p}$. 
A schematic representation is provided in Fig.~\ref{fig:basis}.

The (approximate) $CP$-invariance of the Standard Model imposes that the correlation matrix is symmetric, $\mathbf{C}^I=(\mathbf{C}^I)\trans$, and both spin polarizations are equal, $\mathbf{B}^{I,+}= \mathbf{B}^{I,-}$. Furthermore, at LO QCD, $\rho^{I}(\beta,\hat{k})$ is unpolarized and the $n$-axis is uncorrelated with respect to the remaining directions:
\begin{equation}    \mathbf{B}^{I,\pm}=0,~\mathbf{C}^I=(\mathbf{C}^I)\trans=\left[\begin{array}{ccc}
C^I_{kk} & C^I_{kr} & 0 \\
C^I_{kr} & C^I_{rr} & 0\\
0 & 0 & C^I_{nn}
\end{array}\right].
\end{equation}
It is easily shown that here $\Delta^I>0$ [see Eq. (\ref{eq:DeltaQI})], is a sufficient and also necessary condition for entanglement as the concurrence reads 
\begin{equation}\label{eq:Deltatt}
\mathcal{C}[\rho^I]=\max(\Delta^I,0),~\quad\Delta^I\equiv\frac{-C^I_{nn}+|C^I_{kk}+C^I_{rr}|-1}{2}.
\end{equation}
The concurrence is an even function of $\beta$ and $\cos\Theta$, computed analytically through LO QCD and depicted in Fig.~\ref{fig:Concurrence} for different initial states.

\subsection{$q\bar{q}$ annihilation}\label{subsec:qq}

\begin{figure*}[!tb]
\begin{tabular}{@{}cc@{}}
\stackinset{l}{0pt}{t}{0pt}{(a)}{\includegraphics[width=0.5\textwidth]{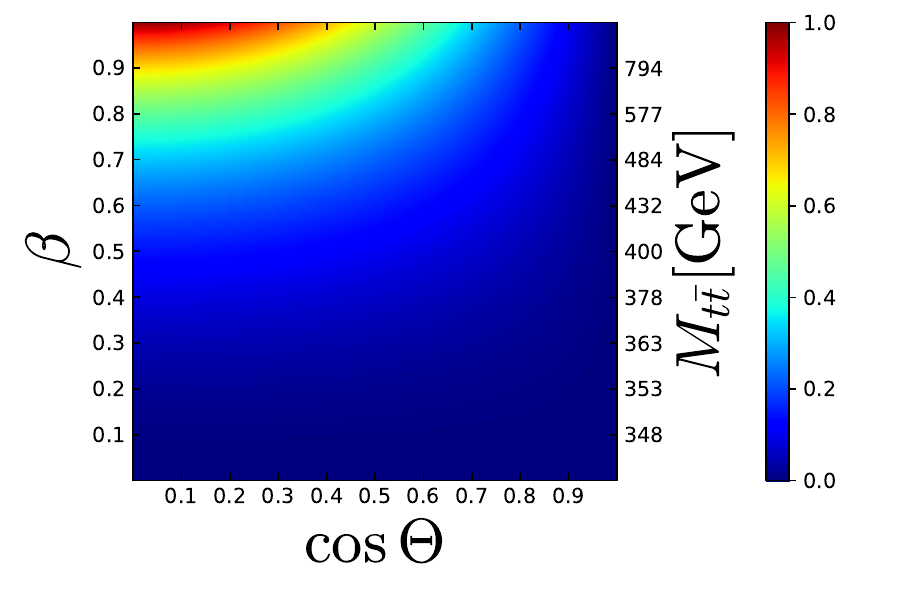}} & \stackinset{l}{0pt}{t}{0pt}{(b)}{\includegraphics[width=0.5\textwidth]{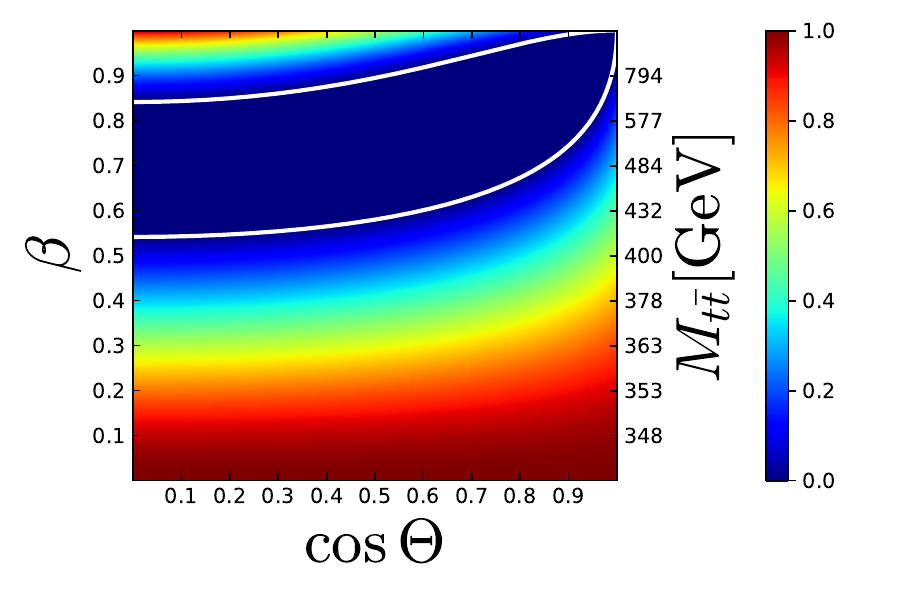}}\\
\stackinset{l}{0pt}{t}{0pt}{(c)}{\includegraphics[width=0.5\textwidth]{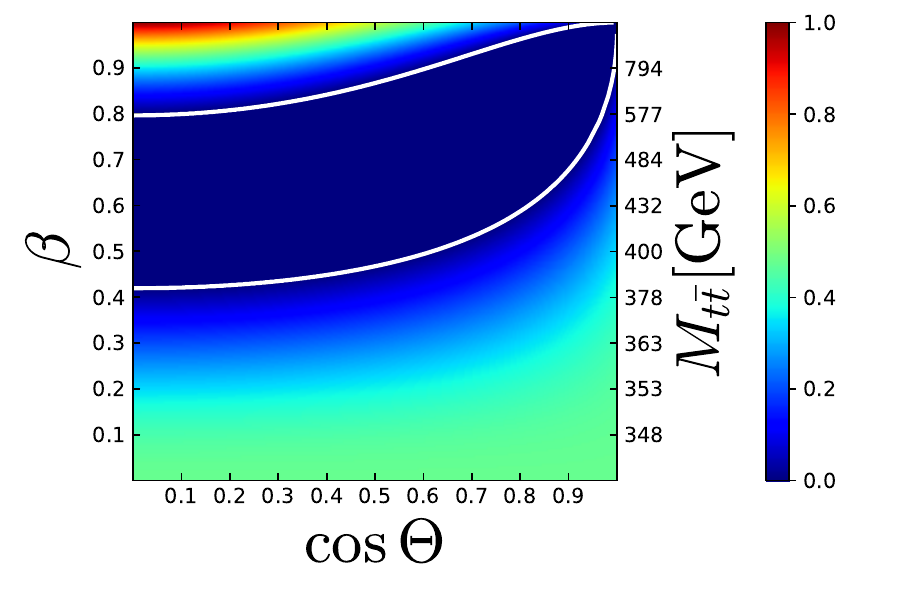}} & \stackinset{l}{0pt}{t}{0pt}{(d)}{\includegraphics[width=0.5\columnwidth]{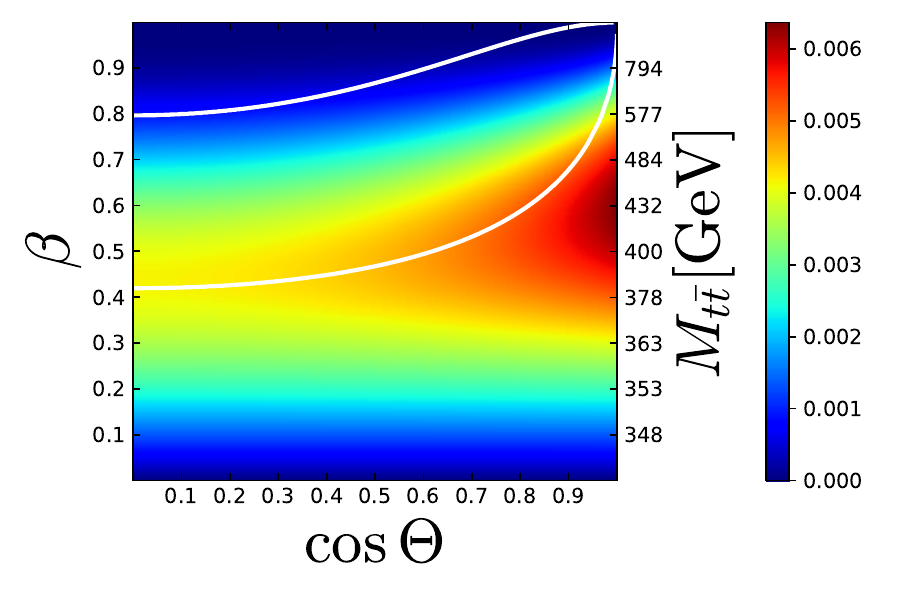}}
\end{tabular}
\caption{Concurrence as a function of $(\beta,\cos\Theta)$. All plots are even functions of $\cos\Theta$. White lines indicate the boundaries between separability and entanglement. (a) $\mathcal{C}[\rho^{q\bar{q}}]$. (b) 
$\mathcal{C}[\rho^{gg}]$. (c) $\mathcal{C}[\rho(\sqrt{s})]$ for the COM energy of LHC Run 2, $\sqrt{s}=13~\textrm{TeV}$. The PDF values used are proportioned by the NNPDF30LO PDF set~\cite{NNPDF:2014otw}. (d) Relative cross section $\frac{1}{\sigma}\frac{d\sigma}{d \Mtt d\cos\Theta}$ in $\textrm{GeV}^{-1}$ for (c).}
\label{fig:Concurrence}
\end{figure*}

For $q\bar{q}$ annihilation, Fig.~\ref{fig:Concurrence}a, the resulting $t\bar{t}$ quantum state $\rho^{q\bar{q}}(\beta,\hat{k})$ is entangled in the bulk of phase space as
\begin{align}
\Delta^{q\bar{q}}=\frac{\beta^2\sin^2\Theta}{2-\beta^2\sin^2\Theta}\geq 0.
\end{align}

The above inequality is saturated only at threshold ($\beta=0$) or for forward/backward production ($\sin\Theta=0$). In both limits, the $t\bar{t}$ spins are aligned along the beam axis in a maximally yet classically correlated mixed state:
\begin{align}\label{eq:qqseparable}
\rho^{q\bar{q}}(0,\hat{k})=\rho^{q\bar{q}}(\beta,\pm\hat{p})=\frac{\ket{\hat{p},\hat{p}}\bra{\hat{p},\hat{p}}+\ket{-\hat{p},-\hat{p}}\bra{-\hat{p},-\hat{p}}}{2}.
\end{align}
Hence, there is little entanglement close to these regions of phase space, as seen in Fig.~\ref{fig:Concurrence}a. In the opposite ultrarelativistic limit of high $p_{\textrm T}$ (i.e., $\beta \simeq 1$ and $\cos\Theta\simeq 0$), the $t\bar{t}$ pair is in a pure spin-triplet state with zero projection along the $n$-axis,
\begin{align}\label{eq:triplet}
\rho^{q\bar{q}}(1,\hat{n}\times\hat{p})=\ket{\Psi_{\infty}}\bra{\Psi_{\infty}},~\ket{\Psi_{\infty}}=\frac{\ket{\hat{n},-\hat{n}}+\ket{-\hat{n},\hat{n}}}{\sqrt{2}}.
\end{align}
This is one of the Bell states~(\ref{eq:BellStates}), maximally entangled, responsible for the strong entanglement signature in the upper left corner of Fig.~\ref{fig:Concurrence}a.

\subsection{$gg$ fusion}\label{subsec:gg}

For $gg$ fusion, Fig.~\ref{fig:Concurrence}b, the resulting $t\bar{t}$ quantum state $\rho^{gg}(\beta,\hat{k})$ is entangled \textit{iff}  
\begin{equation}
    \Delta^{gg}=\frac{\beta^4(1+\sin^4\Theta)-\beta^2(1+\sin^2\Theta)+|1-\beta^2(1+\sin^2\Theta)|}{1+2\beta^2\sin^2\Theta-\beta^4(1+\sin^4\Theta)}>0.
\end{equation}
Two entanglement regions are found: one close to threshold, and another one at high $p_{\textrm T}$, delimited by the critical boundaries (white solid lines in Fig.~\ref{fig:Concurrence}b) 
\begin{align}\label{eq:BetaCriticals}
\beta_{c1}=\sqrt{\frac{1+\sin^2\Theta-\sqrt{2}\sin\Theta}{1+\sin^4\Theta}}\leq
\beta_{c2}&=\frac{1}{(1+\sin^4\Theta)^{\frac{1}{4}}}.
\end{align}

The strong entanglement signature close to threshold results from the fact that the $t\bar{t}$ pair is there in a spin singlet, another Bell state (\ref{eq:BellStates}):
\begin{equation}\label{eq:singletgg}
\rho^{gg}(0,\hat{k})=\ket{\Psi_0}\bra{\Psi_0},~\ket{\Psi_0}=\frac{\ket{\hat{n},-\hat{n}}-\ket{-\hat{n},\hat{n}}}{\sqrt{2}}.
\end{equation}
The other entanglement region at high $p_{\textrm T}$ results from the fact that the \ttbar is there in the same maximally entangled triplet state as for $q\bar{q}$ annihilation, Eq.~(\ref{eq:triplet}).

\subsection{Realistic QCD processes}\label{subsec:EntanglementRealistic}

Because of color confinement, quarks and gluons are never found as free particles in nature, but rather as constituents of hadrons, where they are collectively referred to as partons. Specifically, the composition of a hadron is described by its parton distribution function (PDF), determining the relative contribution of each parton to a given QCD process at a certain energy scale. 
In the particular case of \ttbar production from an initial colliding hadron beam, such as proton-proton collisions at the LHC, the probability of occurrence of each partonic channel $I=q\bar{q},gg$ for given $M_{t\bar{t}}$ and beam COM energy $\sqrt{s}$ is described by the luminosity function $L_{I}(M_{t\bar{t}},\sqrt{s})$~\cite{Bernreuther1998}, computed in terms of the PDFs of the colliding hadrons. 
For instance, the total differential cross section for \ttbar production is the sum of the partonic differential cross sections, weighted by their luminosity functions:
\begin{equation}\label{eq:CrossSectionDifferentialReal}
\frac{\mathrm{d}\sigma}{\mathrm{d}M_{t\bar{t}}\mathrm{d}\Omega}(M_{t\bar{t}},\hat{k},\sqrt{s})=\sum_{I} L_{I}(M_{t\bar{t}},\sqrt{s})\frac{\mathrm{d}\sigma^I}{\mathrm{d}\Omega}(M_{t\bar{t}},\hat{k}).
\end{equation}
Similarly, the total \ttbar quantum state can be written as a statistical mixture of the individual partonic quantum states:
\begin{align}\label{eq:partonicstates}
\nonumber \rho(M_{t\bar{t}},\hat{k},\sqrt{s})&=\sum_{I} w_I(M_{t\bar{t}},\hat{k},\sqrt{s})\rho^{I}(M_{t\bar{t}},\hat{k}),\\
w_I(M_{t\bar{t}},\hat{k},\sqrt{s})&=\frac{L_{I}(M_{t\bar{t}},\sqrt{s})\dfrac{\mathrm{d}\sigma^I}{\mathrm{d}\Omega}(M_{t\bar{t}},\hat{k})}{\dfrac{\mathrm{d}\sigma}{\mathrm{d}\Omega\mathrm{d}M_{t\bar{t}}}(M_{t\bar{t}},\hat{k},\sqrt{s})}. 
\end{align}
The weights $w_I$ represent the probability of production of each individual partonic quantum state $\rho^{I}(M_{t\bar{t}},\hat{k})$, satisfying $w_{q\bar{q}}+w_{gg}=1$. They are given by the probability of occurrence of the partonic channel $I$, $L_I$, multiplied by the probability of producing a $t\bar{t}$ pair through that channel, in turn proportional to the differential partonic cross section of the process. 

The upshot of this discussion is that the elementary $q\bar{q}$ and $gg$ channels are the building blocks of any realistic hadronic \ttbar-production process, where the specific nature of the hadron collision only changes the amount of mixing between them through the PDF-dependent probabilities $w_I(M_{t\bar{t}},\hat{k},\sqrt{s})$. 
From a more general perspective, we can regard the production spin density matrices $R^{I}(M_{t\bar{t}},\hat{k})$ and the luminosities $L_{I}(M_{t\bar{t}},\sqrt{s})$ as HEP inputs, from which we  compute the physical spin density matrices $\rho^I(M_{t\bar{t}},\hat{k})$ and the probabilities $w_I(M_{t\bar{t}},\hat{k},\sqrt{s})$.
Once done, we are simply left with a textbook QI problem involving the convex sum of two-qubit states, to which we can apply the usual techniques of the field.

The concurrence of $\rho(M_{t\bar{t}},\hat{k},\sqrt{s})$ for an initial colliding proton-proton beam at $\sqrt{s}=13~\textrm{TeV}$, corresponding to the LHC setup during Run 2, is depicted in Fig.~\ref{fig:Concurrence}c, while the associated relative differential cross section is shown in Fig.~\ref{fig:Concurrence}d. At the LHC, the $gg$ channel strongly dominates over the $q\bar{q}$ channel, resulting in the resemblance between Figs.~\ref{fig:Concurrence}b and \ref{fig:Concurrence}c.

\section{Experimental implementation}\label{sec:Experimental}

\subsection{Quantum tomography protocol}

In experiments, all magnitudes are averaged over phase space due to the finite resolution and available statistics. Given a certain phase-space region $\Pi$, the corresponding \ttbar spin quantum state is described by the density matrix
\begin{align}\label{eq:TotalQuantumState}
\rho_\Pi(\sqrt{s})&\equiv\int_\Pi\mathrm{d}M_{t\bar{t}}\mathrm{d}\Omega~p(M_{t\bar{t}},\hat{k},\sqrt{s})\rho(M_{t\bar{t}},\hat{k},\sqrt{s}),\\
\nonumber p(M_{t\bar{t}},\hat{k},\sqrt{s})&=\frac{1}{\sigma_\Pi}\frac{\mathrm{d}\sigma}{\mathrm{d}\Omega\mathrm{d}M_{t\bar{t}}}(M_{t\bar{t}},\hat{k},\sqrt{s}),\quad \sigma_\Pi\equiv\int_\Pi\mathrm{d}\Omega\mathrm{d}M_{t\bar{t}}~\dfrac{\mathrm{d}\sigma}{\mathrm{d}\Omega\mathrm{d}M_{t\bar{t}}}(M_{t\bar{t}},\hat{k},\sqrt{s}).
\end{align}
Intuitively, we can understand $\rho_\Pi$ as the convex sum over all possible \ttbar quantum states inside $\Pi$, weighted by the probability $p(M_{t\bar{t}},\hat{k},\sqrt{s})$ of the corresponding production process, proportional to the total differential cross section. 

The physical interest of $\rho_\Pi$ is that its quantum tomography can be experimentally implemented.
This is where the uniqueness of the top quark enters in place: as the most massive fundamental particle of the Standard Model ($m_t\simeq 172.5~\textrm{GeV}$)~\cite{ParticleDataGroup:2026mpi}, it also has a very large width ($\Gamma_t\simeq 1.42~\textrm{GeV}$), resulting in such a short lifetime ($\tau_t\simeq 4.6\cdot10^{-25}~\textrm{s}$) that the top quark decays so quickly that no other process, including hadronisation or spin decorrelation, can affect its spin. 
The information on the top spin is then immaculately imprinted on the flight distribution of its decay products, from where it can be retrieved.
This idea can be also extended to \ttbar decays.
Specifically, for a dileptonic decay $t\bar{t}\to b\nu_\ell \ell^+\bar{b}\bar{\nu}_\ell \ell^-$, the angular differential cross section characterizing the antilepton/lepton directions $\hat{\ell}_{\pm}$ in their respective parent top/antitop rest frames (after boosting from the joint COM frame) reads
\begin{equation}\label{eq:LeptonicCrossSection}
\frac{1}{\sigma}\frac{\mathrm{d}\sigma}{\mathrm{d}\Omega_{+}\mathrm{d}\Omega_{-}}=\frac{1+\mathbf{B}^{+}\cdot\hat{\ell}_{+}-\mathbf{B}^{-}\cdot\hat{\ell}_{-}
-\hat{\ell}_{+}\cdot \mathbf{C} \cdot\hat{\ell}_{-}}{(4\pi)^2},
\end{equation}
$\Omega_{\pm}$ being the solid angles associated to $\hat{\ell}_{\pm}$. 
The vectors $\mathbf{B}^{\pm}$ and the matrix $\mathbf{C}$ are precisely the spin coefficients characterizing the quantum state $\rho_\Pi$. 
Hence, by appropriately fitting the above cross section, which is experimentally accessible, one can measure all spin coefficients, performing in this way the quantum tomography of $\rho_\Pi$.
For such a measurement, the full kinematic reconstruction of each event, including the \ttbar and $\ell^+\ell^-$ momenta, must be performed. 
This not only allows to implement any desired selection on the \ttbar kinematics (such as restricting them to the phase-space region $\Pi$), but it also implies that one can measure the spin coefficients in any arbitrary orthonormal basis, including the helicity basis.

\subsection{Experimental entanglement observation}

We analyze here our original experimental proposal~\cite{Afik2021} leading to the observation of entanglement by the ATLAS (which both of us co-authored) and CMS collaborations using the LHC Run 2 dataset~\cite{ATLAS2024,CMS:2024pts}.
We first stress that the experimental sensitivity is controlled not just by the amount of entanglement, but also by the available statistics. 
From Fig.~\ref{fig:Concurrence}d, we identify the strongest \ttbar signal close to threshold. 
This singles out the spin singlet at threshold as the most accessible target for an entanglement observation. 
Moreover, the invariance under rotations of the spin singlet allows averaging over all possible top directions to increase the signal. 
As a result, the region $\Pi$ in the $(\Mtt,\hat{k})$ phase space is chosen as $(2m_t,\Mtt)\times S^2$, i.e., we impose an upper cut to the top invariant mass at $\Mtt$ while integrating the top flight direction over the whole sphere.

Furthermore, since our target is a spin singlet, we can use the fidelity witness (\ref{eq:DeltaFidelity}), conveniently rewriting it as
\begin{equation}
   \textrm{Tr}[\mathbf{C}]+1<0 \Longleftrightarrow D\equiv \frac{\textrm{Tr}[\mathbf{C}]}{3}<-\frac{1}{3}.
\end{equation}
Thus, $D$ is another entanglement witness, with $D<-1/3$ a sufficient condition for entanglement. 
The key point is that $D$ can be directly extracted from the cross section determining the angular separation between the leptons:
\begin{equation}\label{eq:DMeasure}
\frac{1}{\sigma} \frac{d\sigma}{d\cos\varphi}=\frac{1}{2}(1 - D\cos\varphi),\qquad D = -3\cdot \langle\cos\varphi\rangle,\qquad \cos\varphi=\hat{\ell}_{+}\cdot \hat{\ell}_{-}\,.
\end{equation}
Remarkably, this implies that entanglement can be experimentally signaled by measuring just one parameter. 
Specifically, our proposal suggested to measure $D$, using (\ref{eq:DMeasure}), after applying an upper \Mtt cut.
This experimental scheme was implemented by both ATLAS and CMS using different phase-space selections, resulting in the measured values $D=-0.537\pm 0.019$ and $D=-0.452^{+0.025}_{-0.026}$, respectively. 
Both results observed entanglement with more than $5$ standard deviations ($5\sigma$), which represented the highest-energy observations of entanglement ever.

\section{Conclusions and impact}

Based on our results from Refs.~\cite{Afik2021,Afik2022}, this work reviews how QI can be also studied in high-energy colliders, a unique experimental environment for such purpose due to their fundamental nature and genuinely relativistic character.

From a theoretical point of view, our work provides a neat dictionary translating QI concepts into HEP magnitudes and vice versa, effectively bridging the two fields in a genuinely interdisciplinary way that allows for a rich and continuous exchange between them. 
From a phenomenological point of view, our work shows how to implement a number of QI techniques in colliders, offering novel tools to probe high-energy dynamics, with quantum tomography playing a central role as the knowledge of the quantum state enables the evaluation of all QI magnitudes. 
From an experimental point of view, our work identifies the top quark as a drosophila of a relativistic qubit, proposing specific schemes for entanglement measurements and quantum tomography in \ttbar production.

The subsequent impact of our work has been enormous, around which a rich and expanding field of QI in HEP (QIHEP) has effectively developed. 
A most direct output has been the experimental entanglement observation by the ATLAS and CMS collaborations based on our proposal~\cite{ATLAS2024,CMS:2024pts}, in turn the highest-energy observation of entanglement ever. 
Indeed, the strong sensitivity of our entanglement witness has eventually led to the recent observation of an excess near the \ttbar threshold by ATLAS and CMS~\cite{CMS:2025kzt,ATLAS2026}, which might be explained by non-relativistic QCD effects, sometimes referred to as ``toponium''.
Following our work, other QI concepts such as discord~\cite{Afik2023}, entanglement autodistillation~\cite{Aguilar-Saavedra:2024fig}, steering~\cite{Afik2023}, Bell nonlocality~\cite{Fabbrichesi2021}, magic~\cite{White:2024nuc}, or even enhanced quantum key distribution~\cite{Moreno2026} have been studied in \ttbar production. 
Alternative QIHEP systems were identified after top quarks, including bottom quarks~\cite{Afik2025}, tau leptons~\cite{Fabbrichesi2022}, as well as $W$~\cite{Barr2022} and $Z$~\cite{Aguilar2023} bosons, which provide relativistic qutrit platforms. 
Multipartite entanglement is investigated in many-particle decays~\cite{Sakurai2024}.
Quantum tomography has been adapted to flavor degrees of freedom~\cite{Cheng2026}. 
Another intriguing research line is the application of QI techniques for New Physics searches~\cite{Aoude2022}. 
A recent review can be found in Ref.~\cite{Afik:2025ejh}, co-authored by over 70 researchers (from both QI and HEP, and from both theory and experiment), in which can be regarded as a \textit{quantum manifesto} illustrating the large QIHEP community already gathered.

\section*{Acknowledgments} 

We thank the ICBS committee for selecting our article~\cite{Afik2021} for the FSA award.
We would like to specially thank Fernando Sols for his strong support and wise advice throughout the whole project. 
We also thank J. J. Garc\'ia--Ripoll, Y. Rozen, Y. Shadmi, Y. Soreq, Y. Cohen, Y. Weiss, G. Durieux, G. Sierra, B. Ravina, and T. Kitahara for fruitful discussions and valuable comments.
YA is supported by the National Science Foundation under Grant No.\ PHY-2310094. 
JRMdN acknowledges funding from Spain's MICIU/AEI through Ram\'on y Cajal program (Grant No. RYC2024-050437-I) and through Proyectos de Generaci\'on de Conocimiento (Grant No. PID2025-169103NA-I00).

\bibliographystyle{unsrt}
\bibliography{Enttanglement}

\end{document}